\documentclass[11pt,twoside]{article}
\usepackage{asp2010}

\resetcounters

\def\yu{2005\,YU\ensuremath{_{55}}}
\def\bx{2012\,BX\ensuremath{_{34}}}

\begin{document}

\title{The Fly's Eye Camera System -- an instrument design for large \'etendue time-domain survey}
\author{%
Gergely Cs\'ep\'any$^{1,*}$;
Andr\'as P\'al$^{1,2,**}$;
Kriszti\'an Vida$^{1}$;
Zsolt Reg\'aly$^{1}$;
L\'aszl\'o M\'esz\'aros$^{1,2}$;
Katalin Ol\'ah$^{1}$;
Csaba Kiss$^{1}$;
L\'aszl\'o D\"obrentei$^{1}$;
Attila Jask\'o$^{1}$;
Gy\"orgy Mez\H{o}$^{1}$ and
Ern\H{o} Farkas$^{1}$
\affil{%
$^1$MTA Research Centre for Astronomy and Earth Sciences, 
Konkoly Thege Mikl\'os \'ut 15-17, Budapest, H-1121, Hungary;\\
$^2$Department of Astronomy, Lor\'and E\"otv\"os University, 
P\'azm\'any P\'eter s\'et\'any 1/A, Budapest H-1117, Hungary\\
E-mail addresses: $^*$gcsepany@flyseye.net, $^{**}$apal@flyseye.net
}}

\begin{abstract}
In this paper we briefly summarize the design concepts of the 
{\it Fly's Eye Camera System}, 
a proposed high resolution all-sky monitoring
device which intends to perform high cadence time domain astronomy
in multiple optical passbands while
still accomplish a high \'etendue. Fundings have already been accepted
by the Hungarian Academy of Sciences
in order to design and build a {\it Fly's Eye} device unit. Beyond the
technical details and the actual scientific goals, this paper
also discusses the possibilities and yields of a network operation
involving $\sim10$ sites distributed geographically in 
a nearly homogeneous manner. Currently, we expect 
to finalize the mount assembly -- that performs the sidereal tracking
during the exposures -- until the end of 2012 and to 
have a working prototype with a reduced number of individual cameras
sometimes in the spring or summer of 2013.
\end{abstract}

\section{Introduction}

The key to learn more about the Universe and unveil the physical 
processes beyond various astronomical phenomena is to monitor the 
alterations of observable quantities, such as fluxes in several spectral 
regimes of the sky. Although some of the astrophysical processes have 
their own characteristic timescales, most of the complex systems exhibit 
variations on a broader, currently unexplored temporal spectrum. Hence, 
continuous monitoring of the whole sky will reveal currently 
unknown phenomena and quantify properties of physical events ongoing in 
stellar and planetary systems as well as in their neighborhood. We aim 
to design, build and operate the core of the {\it Fly's Eye Network}: a 
network of geographically distributed large \'etendue and high 
resolution all-sky monitoring devices, extending the currently developed 
single-station operation. This network provides a homogeneous and 
truly full sky investigation of the time domain of astrophysical events 
that covers $\sim6$ magnitudes from the data acquisition cadence of some 
minutes up to the range of the expected operations, i.e. several years. 
The novel hexapod-based arrangement of the camera platform of the 
{\it Fly's Eye} device allows us to install and 
maintain a setup independently of 
the geographical location and without the need of polar adjustment in an 
enclosed dome from which the detectors watch the sky through optical 
windows. Moreover, the construction is highly fault tolerant and lacks 
unique components. All of these enable an easily sustainable 
instrumentation even in harsh environments. Due to its design 
parameters, the resulting network will yield an \'etendue that is 
comparable to that of the Large Synoptic Survey Telescope 
\citep[LSST,][]{ivezic2008}.
Moreover, our concept provides an innovative instrument that is fully 
complementary to the latter, since the {\it Fly's Eye} faint limit is roughly 
the same as the expected saturation brightness of the LSST and the 
employed spectral passbands also match.

\begin{figure}
\begin{center}
\resizebox{60mm}{!}{\includegraphics{O09_f1}}%
\resizebox{60mm}{!}{\includegraphics{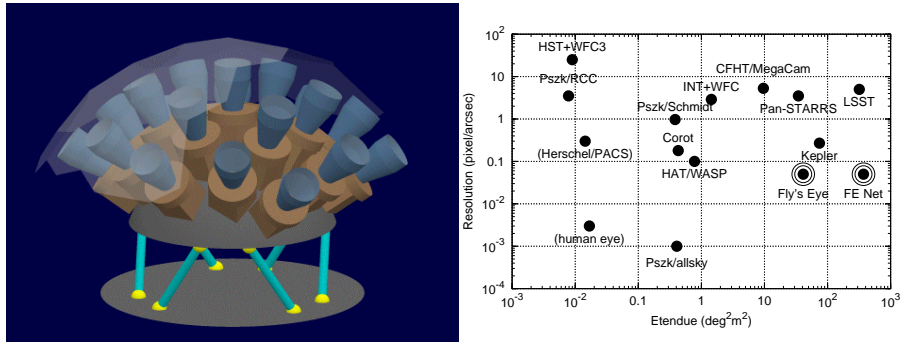}}
\end{center}
\caption{%
{\it Left panel:} a simple on-scale visualization of the camera 
mount and optics. The payload platform with the 19 camera -- lens pairs has
an effective diameter of nearly 1\,m. 
{\it Right panel:} the optical light-collecting phase volume, 
or \'etendue and effective resolution for various known,
mostly optical telescope systems. 
The proposed design of a single {\it Fly's Eye Camera} unit yields a value that 
is comparable to the available 
instruments which have the largest \'etendue: a single unit
of the \emph{Pan-STARRS} telescope(s) and 
the \emph{Kepler} space telescope. A network of nine {\it Fly's Eye} devices
(see labelled as {\it FE Net}) yields an \'etendue comparable to 
the proposed design of \emph{LSST}.}
\label{fig:flyseye}
\end{figure}

\section{Design concepts}

Although ``normal'' all-sky cameras are assembled with fixed optics and 
detectors, the expected per-image exposure times and the required 
resolution\footnote{The resolution is nonetheless smaller than the 
average imaging optical 
telescopes but definitely higher than of the usual all-sky cameras.} requires 
sidereal tracking of the {\it Fly's Eye} optical mount. In order to provide 
a solution for sidereal tracking, we employ a hexapod-based 
design\footnote{\url{http://en.wikipedia.org/wiki/Stewart_platform}} 
for supporting the cameras and optics. These hexapod mounts allow us 
to perform spatial rotations in all of the three rotational degrees
of freedom without any parametric singularity (e.g. gimbal lock). Hence,
this design yields not only a device that is exactly the same 
independently from the actual geographical location but also makes
the otherwise complicated procedure of polar adjustment unnecessary. 
Moreover, hexapods are fault-tolerant mechanisms in this type of application
since the degrees of freedoms associated to the spatial displacements
are not exploited. In other words, even 3 out of the 6 legs can be broken
but the device is still capable to track and compensate the apparent 
rotation of Earth. 

The camera platform itself consists of 19 wide-field cameras, where each
camera lens-system is built from $4{\rm k}\times 4{\rm k}$ detectors
with a pixel size of $9\times9\,{\rm\mu m}$ while the optics are 85mm/f1.2
lenses. This setup yields an effective resolution of $22''/{\rm pixel}$
and an \'etendue of $40\,{\rm m}^2{\rm deg}^2$ and covers the half of the
visible sky. The visualization
of the camera assembly, the hexapod mount and the resulted field-of-view
are displayed in the left panel of Figure~\ref{fig:flyseye}. 
Due to the arrangement
and the large number of individual camera-lens units, the whole design
resembles the compound eyes of insects, hence the name of the project.

The expected data flow rate in continuous operation is about $100$~TB/year.
Taking into account a yearly duty cycle of $0.40$ (for Hungary), the data flow reduces to 40--50~TB/year without compression.
By employing data compression (eg. FPACK) we except to have about 30~TB data to store each year.
Since the most affordable hard disk size nowadays is the 2~TB disk, having 19 disks (one for each camera, totaling in $19\times2=38$~TB) can roughly hold up to one year's data. To secure the data and provide redundancy, the data can be distributed on 24 disks, by adding 5 extra disks to the 19. The 24 disks can be grouped as $6\times4$~disks or $4\times6$~disks, therefore the data can be recovered even if a full unit breaks (in the former setup) or if one disk dies in each unit (in the latter setup). The full redundancy is achieved by using Galois field arithmetics over $GF(2^8)$ or $GF(2^{32})$ and Vandermonde transformation.

\section{Scientific applications}

The main goals of the proposed {\it Fly's Eye} Camera System
and the further Network cover several topics in astrophysics. In the following, 
without attempting to be comprehensive, we list some of these sub-fields of
astrophysics. 

\subsection{Solar System} 
Even with its moderate resolution, 
the {\it Fly's Eye} device is capable to detect meteors and map 
these tracks with an effective resolution of $\sim10$\,m/pixel. Hence,
and due to the large light collecting power of the device, a more
accurate distribution of Solar System dust can be derived. In addition,
for the bright-end of the main-belt asteroid family members, an
unbiased sample will be available for their rotation and shape
properties \citep[see also][]{durech2011}.
These data are essential to understand the aspect
of Solar System dynamics and, more importantly, its evolution. 

Moreover,
nearby flybys of small bodies that are potentially hazardous to
the Earth can be traced (see e.g. the cases of \yu{} and \bx{}). 
Due to the continuous sampling, such information
is also recoverable in an \emph{a posteriori} manner, i.e. if deeper
surveys discover such an object and dynamical calculations confirm
a former approach in the field of view of one or more {\it Fly's Eye} device.

\subsection{Stellar and planetary systems} 
Young stellar objects are complex astrophysical systems and show 
signs of both quasi-periodic and sudden 
transient, eruptive processes. By monitoring their intrinsic variability,
one is able to obtain several constraints regarding to the ongoing processes
\citep{hartmann1996,abraham2009}. 
Persistent monitoring of numerous young stellar 
objects or candidates for young stellar objects will reveal the nature of the 
currently unexplored domains of stellar birth. Since observing campaigns
are organized mostly on daily or yearly basis, the behavior of
such systems is practically unknown on other timescales. 

Stars with magnetic activity show photometric variability on all the
time-domains of the planned instrument, from minutes through hours to
years, just like the Sun does \citep{strassmeier2009}.
Continuous monitoring of the sky opens up a new research area for
active stars: the proposed device allows us to obtain good flare
statistics since flares occur on minutes-hours timescale,
and to monitor starspot evolution, differential rotation 
and activity cycles of the same star
\citep[see e.g.][]{hartman2011,walkowicz2011,olah2009}.

Observations of eclipsing binaries provide direct measurements of
stellar masses and radii that are essential to understand their evolution
and even the basic physical processes ongoing in the stellar cores
\citep{latham2009}. Similarly to eclipsing binaries, transiting 
extrasolar planets are also expected to be discovered by the
{\it Fly's Eye Network}, since instruments with nearly similar types
of optics are found to be rather 
efficient \citep{pollacco2004,bakos2004,pepper2007}.

\subsection{In the extragalactic environment}
Continuous monitoring of brighter supernovae in nearby galaxies
yield valuable data that can be exploited by combining
other kind of measurements. The {\it Fly's Eye} camera is capable to observe
the brightest supernovae directly even up to a month
during their peak brightness \citep[see e.g.][]{vinko2012}
and by combining images, it is possible to go even deeper in brightness
using more sophisticated ways of photometric techniques. 

\acknowledgements 
The {\it Fly's Eye} project is supported by the Hungarian Academy of
Sciences via the ``Lend\"ulet'' grant LP2012-31/2012. Additional support
is also received via the ESA grant PECS~98073.

\bibliography{O09}

\end{document}